\documentclass[sigconf]{acmart}

\copyrightyear{2020} 
\acmYear{2020} 
\setcopyright{acmcopyright}\acmConference[CIKM '20]{Proceedings of the 29th ACM International Conference on Information and Knowledge Management}{October 19--23, 2020}{Virtual Event, Ireland}
\acmBooktitle{Proceedings of the 29th ACM International Conference on Information and Knowledge Management (CIKM '20), October 19--23, 2020, Virtual Event, Ireland}
\acmPrice{15.00}
\acmDOI{10.1145/3340531.3412136}
\acmISBN{978-1-4503-6859-9/20/10}
\usepackage{color}
\usepackage[utf8]{inputenc}
\usepackage{longtable}
\usepackage{threeparttable}
\usepackage{xspace}
\usepackage{appendix}
\usepackage{booktabs}
\usepackage[utf8]{inputenc}
\usepackage{subfig}
\usepackage{textcomp}
\usepackage{amsmath, amssymb, amsthm}
\usepackage{enumitem}
\usepackage{algorithm}
\usepackage[noend]{algpseudocode}
\algtext*{EndProcedure}
\usepackage{xcolor,colortbl}
\usepackage{footnote}
\usepackage{url}
\usepackage{tabularx}
\usepackage{multicol}
\usepackage{multirow}
\usepackage{pifont}
\usepackage{bm}
\usepackage{listings}
\usepackage{wrapfig}
\usepackage{relsize}
\usepackage{mathtools}
\usepackage{bbm}
\usepackage{blindtext}
\usepackage{bm}
\usepackage{balance}
\usepackage{mdwlist}
\setlist[enumerate]{leftmargin=*}
\setlist[itemize]{leftmargin=*}

\usepackage{wasysym}
\usepackage{amssymb}
\usepackage{pifont}

\newcommand{\adj}{\mathbf{A}}
\newcommand{\graph}{G}
\newcommand{\vertexSet}{\mathcal{V}}

\newcommand{\embone}{\mathbf{Y}_1}

\newcommand{\embtwo}{\mathbf{Y}_2}
\newcommand{\embonet}{\mathbf{Y}_{1_t}}
\newcommand{\embtwot}{\mathbf{Y}_{2_t}}

\newcommand{\matA}{\mathbf{A}}

\newcommand{\matP}{\mathbf{P}}
\newcommand{\matQ}{\mathbf{Q}}
\newcommand{\matG}{\mathbf{G}}
\newcommand{\matU}{\mathbf{U}}
\newcommand{\matV}{\mathbf{V}}

\newcommand{\matY}{\mathbf{Y}}

\newcommand{\matchMat}{\mathbf{P}}
\newcommand{\perm}{\overline{\mathbf{P}}}

\newcommand{\lr}{\eta}
\newcommand{\reg}{\lambda}
\newcommand{\degmax}{\Delta^*}

\newcommand{\method}{\text{CONE-Align}\xspace}

\newcounter{theorem}

\author{Xiyuan Chen}
\affiliation{\institution{University of Michigan, Ann Arbor}}
\email{shinech@umich.edu}

\author{Mark Heimann}
\affiliation{\institution{University of Michigan, Ann Arbor}}
\email{mheimann@umich.edu}

\author{Fatemeh Vahedian}
\affiliation{\institution{University of Michigan, Ann Arbor}}
\email{vfatemeh@umich.edu}

\author{Danai Koutra}
\affiliation{\institution{University of Michigan, Ann Arbor}}
\email{dkoutra@umich.edu}

\begin{document}
\fancyhead{}

\begin{abstract}

Network alignment, the process of finding correspondences between nodes in different graphs, has many scientific and industrial applications. Existing unsupervised network alignment methods find suboptimal alignments that break up node neighborhoods, i.e. do not preserve \emph{matched neighborhood consistency}. To improve this, we propose \method, which models intra-network proximity with node embeddings and uses them to match nodes across networks after aligning the embedding subspaces. Experiments on diverse, challenging datasets show that \method is robust and obtains 19.25\% greater accuracy on average than the best-performing state-of-the-art graph alignment algorithm in highly noisy settings.
\end{abstract}

\title{CONE-Align: Consistent Network Alignment with Proximity-Preserving Node Embedding} 
\maketitle

\section{Introduction}
\label{sec:intro}
Graphs or networks are ubiquitous structures for representing complex interconnections between entities. One important graph-based data mining task is network alignment: finding correspondences between nodes in different graphs.  
This task has diverse, important applications, such as 
recommendation on social networks, protein-protein interaction analysis, and database schema matching~\cite{kazemi2016network}. 

This work is inspired by a common limitation of network alignment methods.  We find that many unsupervised graph alignment approaches (e.g., FINAL~\cite{final}, NetAlign~\cite{netalign}, REGAL~\cite{regal}) fail to achieve \emph{matched neighborhood consistency}: nodes that are close in one graph are often not matched to nodes that are close in the other graph.  For example, REGAL~\cite{regal} matches nodes using \emph{node embeddings} capturing each node's structural role in the network. However, neighboring nodes may not have similar structural roles, resulting in very different embeddings that may be matched far apart in the other graph, 
violating matched neighborhood consistency.  

To solve this problem, we propose \method for \textbf{\underline{CON}}sistent \textbf{\underline{E}}mbedding-based Network \textbf{\underline{Align}}ment.
We learn similar node embeddings for neighboring nodes in each graph using well-known proximity-preserving node embedding methods.
However, because nodes are not in proximity across graphs, these methods are transductive, and nodes in different graphs will be embedded into different subspaces. Therefore, we align the graphs' embedding subspaces, and then  
we can match the nodes using embedding similarity. Since neighboring nodes in each graph will have similar embeddings, they will be matched to similar parts of the other graph. Thus, we have the best of both worlds with \method: matched neighborhood consistency and cross-graph comparability.  

Our contributions can be summarized as follows:
\begin{itemize}
     \item \textbf{Insights for Network Alignment}: We define the principle of \emph{matched neighborhood consistency}, which motivates us to use node embedding methods with a different kind of objective than what has been used for unsupervised network alignment.
    \item \textbf{Principled New Method}: We propose \method for unsupervised network alignment, which makes embedding subspaces for different graphs comparable, analogous to machine translation using monolingual word embeddings. 
    \item \textbf{Rigorous Experiments}: On challenging datasets, we show that \method outperforms the strongest baseline by 19.25\% on average in accuracy, as it better preserves matched neighborhood consistency. Our code is available at \url{https://github.com/GemsLab/CONE-Align}. 
\end{itemize}

\section{Related Work}
\label{sec:related}

\noindent \textbf{Node Embeddings.} Node embeddings are latent feature vectors modeling relationships between nodes and/or structural characteristics, learned with various shallow and deep architectures and used for many graph mining tasks~\cite{gemsurvey}.  Most embedding objectives model \textbf{proximity} within a single graph: nearby nodes (e.g. neighbors sharing an edge or nodes with mutual neighbors) have similar features. For example, DeepWalk~\cite{deepwalk} and node2vec~\cite{node2vec} perform random walks starting at each node to sample context nodes, using a shallow neural architecture to embed nodes similarly to their context.  This process implicitly factorizes a node pointwise mutual information matrix, which NetMF~\cite{netmf} instead directly factorizes. 

In contrast, \textbf{structural} embedding methods capture a node's structural role independent of its proximity to specific nodes; this independence makes embeddings comparable across graphs~\cite{regal}. For example, struc2vec ~\cite{struc2vec} resembles DeepWalk and node2vec but performs random walks on an auxiliary structural similarity graph. 
xNetMF embeddings~\cite{regal} capture local neighborhood connectivity.  For more on the distinction between structural and proximity-preserving node embeddings, we refer the reader to~\cite{rossi2019community}.

\vspace{0.1cm}
\noindent \textbf{Network Alignment.}   
Classic graph alignment approaches often formulate an \textbf{optimization-based assignment} problem. For example, the message-passing algorithm NetAlign~\cite{netalign} tries to preserve ``complete squares'' by matching two nodes sharing an edge in one graph to counterparts sharing an edge in the other graph.  FINAL~\cite{final} optimizes a topological consistency objective which may be augmented with node and edge attribute information.  
Our approach is initialized by the solution to a classic convex optimization formulation~\cite{gold1996graduated}, but to improve the accuracy, we turn to a different class of methods: those that \textbf{compare node embeddings}.

REGAL~\cite{regal} matches xNetMF structural embeddings that are comparable across networks. Subsequent work ~\cite{du2019joint} models intra-network proximity via link prediction, but its cross-network comparison is also based on structural similarity.  To use transductive proximity-preserving embedding objectives, workarounds include connecting the two graphs with ground truth ``seed'' alignments~\cite{liu2016aligning} if any are known, or using adversarial training techniques in machine translation~\cite{lample2018word} as in another recent work~\cite{dana}. 

\section{Preliminaries}
\label{sec:preliminaries}
\noindent \textbf{Graphs and Embeddings.} We consider two graphs $\graph_1$ and $\graph_2$
with nodesets $\vertexSet_1,\vertexSet_2$ 
and adjacency matrices $\matA_1, \matA_2$ containing edges between nodes. As in~\cite{regal}, for simplicity, we assume that both graphs have $n$ nodes (if not, we can add singleton nodes to one graph).  For each graph $\graph_i$, we can create an $n \times d$ matrix $\matY_i$ of $d$-dimensional node embeddings.

\begin{wrapfigure}{r}{0.2\textwidth}
    \centering
    \vspace{-0.5cm}
    \includegraphics[width=0.2\textwidth]{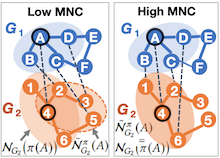}
    \caption{Partial alignments between $\graph_1$ \& $\graph_2$: varying matched neighborhood consistency for node $A$ in $\graph_1$ and its counterpart in $\graph_2$, node $4$.} 
    \label{fig:mnc-example}
\vspace{-0.4cm}
\end{wrapfigure}

\vspace{0.1cm}
\noindent \textbf{Alignment.}
An alignment between the nodes of two graphs is a function $\pi: \vertexSet_1 \rightarrow \vertexSet_2$, or alternatively a matrix $\matchMat$, where $p_{ij}$ is the (real-valued or binary) similarity between node $i$ in $G_1$ and node $j$ in $G_2$. 
A mapping $\pi$ can be found from $\matchMat$, e.g. greedy alignment $\pi(i) = \arg\max_j p_{ij}$.  

\vspace{0.1cm}
\noindent \textbf{Neighborhood.}  
Let $\mathcal{N}_{\graph_1}(i)$ be the neighbors of node $i$ in $\graph_1$, i.e., nodes that share an edge with $i$.  
We define node $i$'s ``\textbf{mapped neighborhood}'' in $\graph_2$ as the 
set of nodes onto which $\pi$ maps $i$'s neighbors: $\tilde{\mathcal{N}}_{\graph_2}^\pi(i) = \{j \in \vertexSet_2 : \exists k \in \mathcal{N}_{\graph_1}(i) \text{ s.t. } \pi(k) = j\}$. 
Also, we denote the neighbors of node $i$'s counterpart $\pi(i)$ as $\mathcal{N}_{\graph_2}\big(\pi(i)\big)$.
We define the \textbf{matched neighborhood consistency (MNC)} of node $i$ in $G_1$ and $j$ in $G_2$ as the Jaccard similarity of the two sets (visualized for a toy example in Fig.~\ref{fig:mnc-example}):
\begin{equation}
\small
\label{eq:jaccard}
MNC(i,j) = \frac{|\tilde{\mathcal{N}}_{\graph_2}^\pi(i) \cap \mathcal{N}_{\graph_2}(j)|}{|\tilde{\mathcal{N}}_{\graph_2}^\pi(i) \cup \mathcal{N}_{\graph_2}(j)|}\\
\end{equation}


\vspace{-0.1cm}
\noindent \textbf{Problem Statement.}
Given two graphs $\graph_1$ and $\graph_2$ with meaningful node correspondences, but none known \textit{a priori}, 
we seek to recover their alignment $\pi$ 
while achieving high MNC.

\section{Method}
\label{sec:method}
\begin{figure}[t!]
    \centering
    \includegraphics[width=0.45\textwidth]{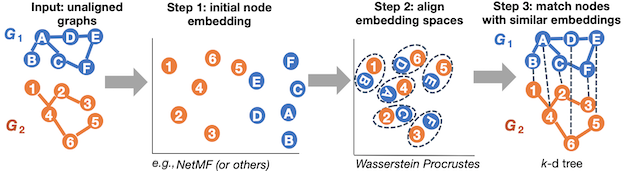}
    \caption{Overview of \method. Given two graphs $\graph_1$ and $\graph_2$, we first use node embedding to model intra-graph node proximity.  Second, we align the embedding spaces 
    for cross-graph comparability.  Third, we match each node in $\graph_1$ to the node in $\graph_2$ with the most similar embedding.}
    \label{fig:method}
    \vspace{-0.5cm}
\end{figure}

 We detail \method (Fig.~\ref{fig:method}, with pseudocode in Alg.~\ref{alg:method}), our proposed method using node embeddings to respect matched neighborhood consistency and to identify cross-graph node similarities.

\subsection{Step 1: Node Embedding}
In Step 1, we obtain normalized node embeddings $\embone, \embtwo \in \mathbb{R}^{n \times d}$ separately per input graph. \method is a framework with which we can use many popular embedding methods, graph neural networks, etc.~\cite{gemsurvey}, \emph{even though} they may be designed for a single network. We only need the embeddings to preserve intra-graph node proximity, i.e. neighboring nodes in each graph have similar embeddings and will be mapped close by when using embedding similarity.  This preserves MNC \emph{robustly}: even when nodes are not neighbors due to missing edges~\cite{du2019joint}, many node embedding algorithms can preserve any higher-order proximities they share.

\subsection{Step 2: Embedding Space Alignment}
Due to the invariance of the embedding objective, the two graphs' node embeddings $\embone$ and $\embtwo$ may be translated, rotated, or rescaled relative to each other. Thus, to compare them, in Step 2, we align the embedding subspaces.  Inspired by unsupervised word translation~\cite{wassersteinprocrustes}, we jointly solve two optimization problems:

\vspace{0.2cm}
\noindent \textbf{Procrustes.} If \emph{node correspondences} were known, we could find a linear embedding transformation $\matQ$ from the set of orthogonal matrices $\mathcal{O}^{d}$. $\matQ$ aligns the columns of the node embedding matrices, i.e. the embedding spaces. It can be obtained by solving an orthogonal Procrustes problem: 
\begin{equation}
\label{eq:orthproc}
\min_{\matQ\in \mathcal{O}^{d}}||\embone\matQ - \embtwo||_2^2   \text{\quad({\it column permutation})}
\end{equation}
Its solution is $\matQ^*=\matU\matV^\top$, where $\matU\mathbf{\Sigma}\matV^T$ is the SVD of $\embone^\top \embtwo$~\cite{orthogonal}.

\vspace{0.2cm}
\noindent \textbf{Wasserstein.} If the \emph{embedding space transformation} were known, we could solve for the optimal node correspondence $\matP$ from the set of permutation matrices $\mathcal{P}^{n}$. $\matP$ aligns the rows of the node embedding matrices, i.e. the nodes. It can be obtained using the Sinkhorn algorithm~\cite{cuturi2013sinkhorn} to minimize the squared Wasserstein distance:
\begin{equation}
\label{eq:wasserstein}
\min_{\matP\in \mathcal{P}^{n}}||\embone-\matP\embtwo||_2^2 \text{\quad({\it row permutation})}
\end{equation}

\noindent \textbf{Wasserstein Procrustes.} As we know neither the \emph{correspondences} nor the \emph{transformation}, we combine the problems:
\begin{equation}
\label{eq:wp}
\min_{\matQ \in \mathcal{O}^{d}}\min_{\matP \in \mathcal{P}^{n}}||\embone\matQ - \matP\embtwo||_2^2
 \end{equation}
We equivalently solve $\max_{\matP\in \mathcal{P}^{n}}\max_{\matQ\in \mathcal{O}^{d}} \text{trace}(\matQ^\top \embone^\top \matP \embtwo)$ 
with a stochastic optimization scheme~\cite{wassersteinprocrustes}, alternating between the Wasserstein and Procrustes problems. For $T$ iterations, 
we use the current embedding transformation $\matQ$ to find a matching $\matP_t$ for minibatches $\embonet, \embtwot$ of $b$ embeddings each, using Sinkhorn~\cite{cuturi2013sinkhorn} with regularization parameter $\reg$. We then use the gradient of the Wasserstein Procrustes distance $||\embonet\matQ - \matP_{t}\embtwot||_2^2$, evaluated on the minibatches $\embonet, \embtwot$, 
to update $\matQ$ with gradient descent (Alg.~\ref{alg:WP}).  

\vspace{0.2cm}
\noindent \textbf{Convex Initialization.}
To initialize the above \emph{nonconvex} procedure, we turn to a classic convex graph matching formulation~\cite{gold1996graduated}:  

\begin{equation}
\label{eq:init}
\min_{\matP\in \mathcal{B}^{n}}||(\adj_1 \matP - \matP \adj_2)||_2^2
\end{equation}
where $\mathcal{B}^{n}$ is the convex hull of $\mathcal{P}^{n}$. 
We can find the global minimizer $\matP^{*}$ with the Frank-Wolfe algorithm ~\cite{frank1956algorithm} for $n_0$ iterations and Sinkhorn~\cite{cuturi2013sinkhorn} with regularization parameter $\reg_0$.  
Using $\embone$ and $\matP^{*}\embtwo$, an initial $\matQ$ can be generated with orthogonal Procrustes (Eq.~\eqref{eq:orthproc}).

\vspace{0.2cm}
\noindent \textbf{Complexity Considerations}.  Our subspace alignment procedure (Alg.~\ref{alg:WP}) uses SVD and Sinkhorn's algorithm~\cite{cuturi2013sinkhorn} on the full data.  Although these algorithms have quadratic time complexity, 
recent superlinear approximations~\cite{altschuler2019massively,allen2016lazysvd} can further scale up \method.

\subsection{Step 3: Matching Nodes with Embeddings}
{\small
\setlength{\textfloatsep}{2pt}
\begin{algorithm}[t!]
\caption{\text{Align Embeddings}\xspace($\embone$, $\embtwo$, $\adj_1$, $\adj_2$)}
\label{alg:WP}
\begin{algorithmic}[1]
\State \textbf{Input:}  node embeddings $\embone$, $\embtwo$, adjacency matrices $\adj_1, \adj_2$

\vspace{0.15cm}
\Statex \textbf{\sf \textcolor{gray}{/* Convex Initialization */}}
\State $\matP^{*} = \arg\min_{\matP\in \mathcal{B}^{n}}||(\adj_1 \matP - \matP \adj_2)||_2^2$
\hfill \textcolor{gray}{\Comment{Initial node correspondences:}} 
\Statex  \textcolor{gray}{\Comment{based on Franke-Wolfe~\cite{frank1956algorithm} and Sinkhorn~\cite{cuturi2013sinkhorn}}}  
\State $\matU\mathbf{\Sigma}\matV^\top = \text{SVD}(\embone^\top \matP^{*}\embtwo$)

\State $\matQ=\matU\matV^\top$
\textcolor{gray}{\Comment{Compute initial embedding space transformation}}

\vspace{0.1cm}
\Statex \textbf{\sf \textcolor{gray}{/* Stochastic Alternating Optimization */}}
\For{$t = 1 \to T$}
\textcolor{gray}{\Comment{T: \# iter (e.g., 50)}}
        \State $\matP_{t} =  \arg\max_{\matP\in \mathcal{P}^{b}}\text{trace}(\matQ^\top \embonet^\top \matP \embtwot)$ \textcolor{gray}{\Comment{Via Sinkhorn, compute}}
        \Statex \textcolor{gray}{\Comment{optimal matching for size-$b$ (e.g., 10) minibatches $\embonet$, $\embtwot$}}
        \State  $\matG_{t} = -2\embonet^\top \matP_{t}\embtwot$ \textcolor{gray}{\Comment{Compute gradient  of WP distance wrt $\matQ$}}
        \State $\matU\mathbf{\Sigma}\matV^\top = \text{SVD}(\matQ - \lr\matG_{t})$ \textcolor{gray}{\Comment{Update orthogonal transform. matrix}}
        \State $\matQ=\matU\matV^\top$
        \textcolor{gray}{\Comment{$\lr$: learning rate (e.g., 1.0)}}
\EndFor
\State \Return{$\matQ$} \textcolor{gray}{\Comment{orthogonal transformation $\matQ$}} 
\end{algorithmic}
\end{algorithm}
}

{\small
\setlength{\textfloatsep}{1pt}
\begin{algorithm}[t!]
\caption{\method($\adj_1, \adj_2$)}
\label{alg:method}
\begin{algorithmic}[1]
\State \textbf{Input:} adjacency matrices $\adj_1, \adj_2$


\vspace{0.15cm}
\Statex \textbf{\sf \textcolor{gray}{/* STEP 1. Model Intra-Network Proximities with Embeddings */}}
\State $\embone = \text{proximity-emb-method}(\adj_1)$, $\embtwo = \text{proximity-emb-method}(\adj_2)$
\State $\embone = \frac{\embone}{||\embone||_2}$, $\embtwo = \frac{\embtwo}{||\embtwo||_2}$\textcolor{gray}{\Comment{Normalize the node embeddings}} 

\vspace{0.1cm}
\Statex \textbf{\sf \textcolor{gray}{/* STEP 2. Align Embedding Spaces for Cross-Graph Comparability */}}
\State $\matQ = \text{Align Embeddings}\xspace(\embone, \embtwo, \adj_1, \adj_2)$ 

\vspace{0.1cm}
\Statex \textbf{\sf \textcolor{gray}{/* STEP 3. Match Nodes with Similar Embeddings */}}
\State $\embone = \embone\matQ$\textcolor{gray}{\Comment{Align embedding spaces and greedily match nodes with}}
\State $\matP = \text{QueryKDTree}(\embone, \embtwo)$\textcolor{gray}{\Comment{sim.\ embeddings via $k$-d tree (NN search)}}
\State \Return{$\matP$} \textcolor{gray}{\Comment{permutation matrix with aligned nodes across input graphs}}
\end{algorithmic}
\end{algorithm}
}
After aligning the embeddings 
with the final transformation $\matQ$, in Step 3, we match each node in $\graph_1$ to its nearest neighbor in $\graph_2$ based on Euclidean distance. We could use scaling corrections to mitigate ``hubness'' whereby many nodes have the same nearest neighbor~\cite{wassersteinprocrustes}, but we found no need.
Following~\cite{regal}, we use a $k$-d tree for fast nearest neighbor search between $\embone \matQ$ and $\embtwo$.  

\section{Experiments}
\label{sec:experiments}
\begin{figure*}[t!]
	\centering
		
    \subfloat[Arenas\label{arenas}]{%
      \includegraphics[width=0.23\textwidth]{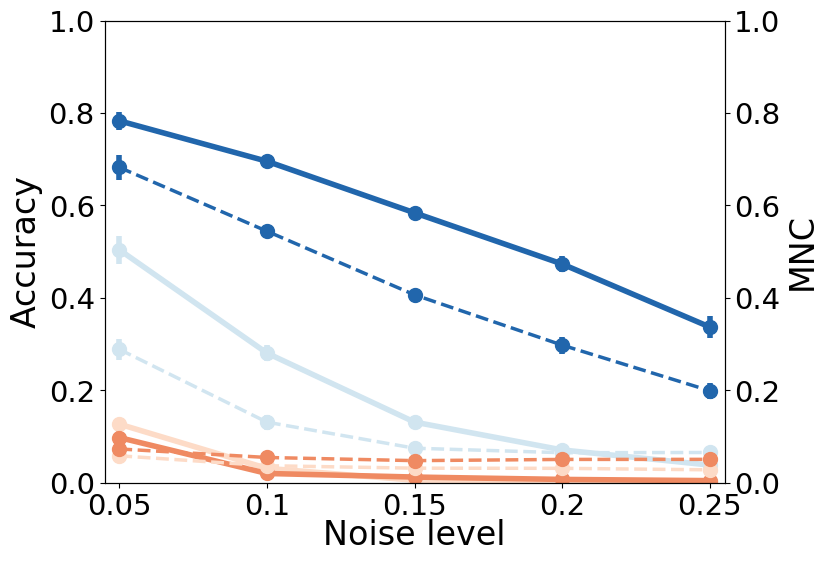}
    }
    ~
    \subfloat[Hamsterster\label{hamster}]{%
      \includegraphics[width=0.23\textwidth]{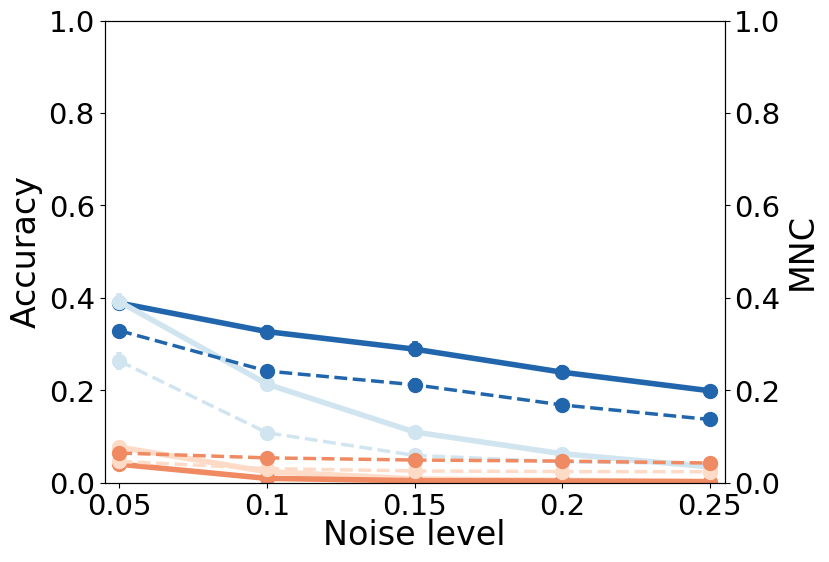}
    }
    ~
    \subfloat[PPI\label{ppi}]{%
      \includegraphics[width=0.23\textwidth]{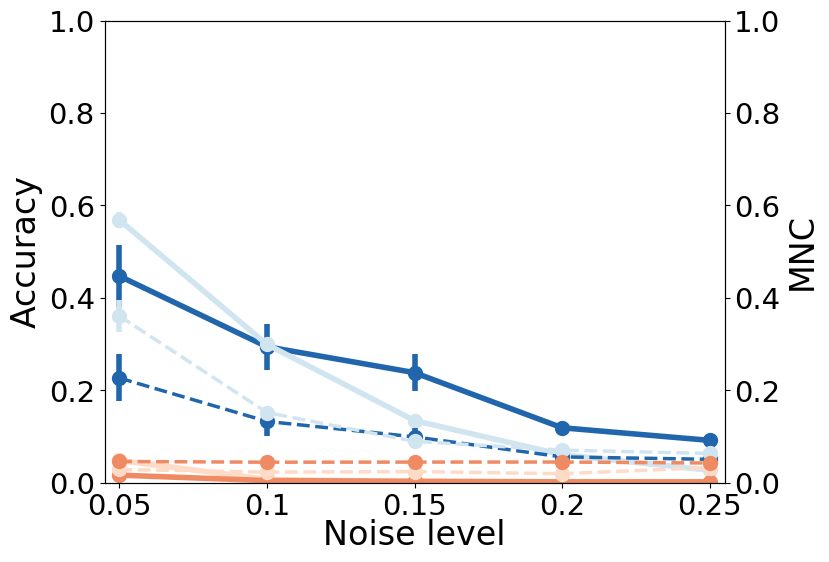}
    }
    ~
    \subfloat[Facebook
    \label{fb}]{\includegraphics[width=0.23\textwidth]{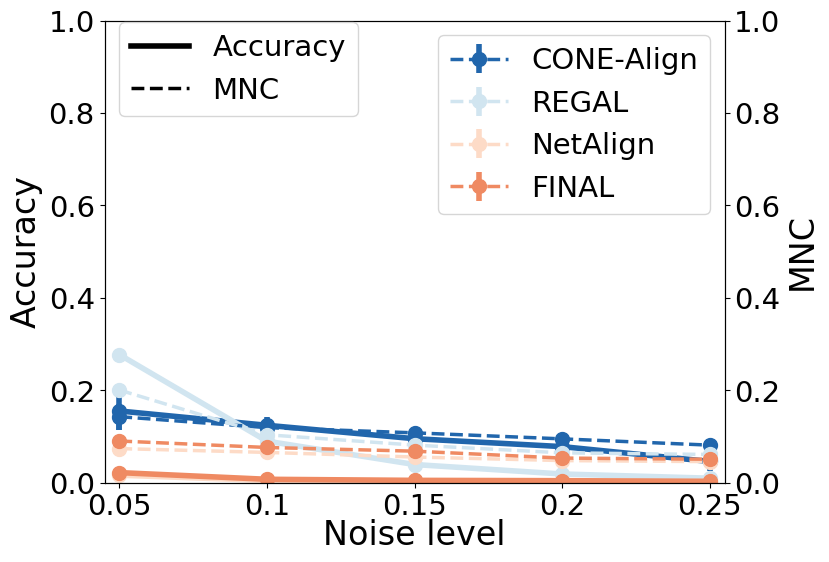}
    }
    \vspace{-0.35cm}
	\centering
\caption{Average accuracy (solid lines) and MNC (dashed lines), with standard deviation in error bars, vs.\ different noise levels.  \method significantly outperforms baselines and better preserves MNC across datasets, particularly as noise increases.}  
    \label{fig:simulated-acc}
    \vspace{-0.4cm}
\end{figure*}

In this section, we analyze \method's accuracy and matched neighborhood consistency in network alignment.

\noindent\textbf{Configuration of \method.}
\label{sec:config} We use NetMF~\cite{netmf} node embeddings, which we find obtain higher accuracy than the related DeepWalk and node2vec~\cite{deepwalk,node2vec}, possibly because the latter use random walks that increase variance~\cite{regal}. 
We use default values~\cite{netmf}, approximating the normalized graph Laplacian with 256 eigenpairs, and set embedding dimension $d=128$, context window size $w=10$, and $\alpha=1$ negative samples~\cite{netmf}.
For the subspace alignment, we use parameters which yield good accuracy and speed: ${n_0} = 10$ iterations and regularization $\reg_0 = 1.0$ for the initial convex matching, and 
$T = 50$ iterations of Wasserstein Procrustes optimization 
with batch size $b = 10$, learning rate $\lr = 1.0$, and regularization $\reg = 0.05$. 

\label{subsec:exp-setup}

     
     
     

\vspace{0.2cm}
\noindent \textbf{Data.} 
Following prior works~\cite{regal,final,dana}, we simulate a network alignment scenario with known ground truth: a graph with adjacency matrix $\adj$ is aligned to a noisy permuted copy $\adj^*$.  We generate a random permutation matrix $\perm$ and set $\adj^* = \perm \adj \perm^\top$; we then randomly remove edges from $\adj^*$ with probability $p \in [0.05, 0.10, 0.15, 0.20, 0.25]$.  We perform this procedure on graphs representing various phenomena as shown in Table~\ref{tab:datasets}.

\begin{table}[h!]
\vspace{-0.3cm}
\centering
\caption{Description of the datasets used.}
\label{tab:datasets}
\vspace{-0.2cm}
{\small 
\begin{tabular}{l r@{\hspace{1.2em}}  r@{\hspace{4pt}} @{\hspace{5pt}}r l}
\toprule
   \textbf{Name} & \textbf{Nodes} & \textbf{Edges} &  \textbf{Description}  \\
\midrule
     Arenas \cite{koblenz} &  1\,133      & 5\,451    & communication network \\ 
     
     Hamsterster \cite{koblenz} & 2\,426 & 16\,613 & social network \\
     
     PPI \cite{ppi} & 3\,890     & 76\,584    &  protein-protein interaction \\
     
     Facebook \cite{snapnets} & 4\,039 & 88\,234 & social network \\ 

\bottomrule
\end{tabular}
}
\vspace{-0.2cm}
\end{table}

\noindent \textbf{Baselines.} 
\label{sec:baselineset}
Our baselines are unsupervised methods using diverse techniques (belief propagation, spectral methods, and embeddings): \textbf{(1)~NetAlign}~\cite{netalign}, \textbf{(2) FINAL}~\cite{final}, and \textbf{(3) REGAL}~\cite{regal}. We configure each method following the literature. NetAlign and FINAL require a matrix of prior alignment information, for which we take the top $k=\lfloor \log_2 n\rfloor$ most similar nodes by degree for each node~\cite{regal,dana}.  
For REGAL we use recommended embedding dimension $\lfloor 10 \log_2(2n) \rfloor$, maximum neighbor distance $2$ with discount factor $\alpha = 0.1$, and resolution parameter $\gamma_{\text{struc}} = 1$~\cite{regal}. 

\subsection{Alignment Performance}
\label{sec:acc}
\subsubsection{Evaluation.} We measure \textbf{alignment accuracy}, or the proportion of correctly aligned nodes, as well as the average \textbf{matched neighborhood consistency (MNC)} using Eq.~\eqref{eq:jaccard} across all nodes.
 
\subsubsection{Results}

In Fig.~\ref{fig:simulated-acc}, we report average and standard deviation for each metric over five trials for each experimental setting.

\textbf{\method outperforms baselines.} We study 
$5\times$ higher noise levels than prior work~\cite{regal}; in this challenging setting, NetAlign and FINAL achieve <10\% accuracy. 
REGAL is most accurate on the PPI and Facebook networks with low noise; in these denser graphs, many nodes share first-order proximities and are hard to distinguish.  However, \method outperforms it above 10\% noise.

\textbf{\method is more robust to noise.}  
\method's accuracy declines less sharply than REGAL as noise increases, and it is the only method to measurably align any datasets at 25\% noise.

\textbf{Accuracy and MNC are closely related.} They trend similarly, and more accurate methods (esp. \method) have higher MNC. 

\textit{Runtime.}
\method's average runtime per dataset ranges from 5 sec to 4 min: slower than the famously scalable methods NetAlign and REGAL, but at least twice as fast as FINAL.  

\subsection{Matched Neighborhood Consistency}

\begin{figure}[t!]
    \vspace{-0.3cm}
	\centering
	
	\subfloat[MNC with \method\label{fig:easy-distafter}]{\includegraphics[width=0.22\textwidth]{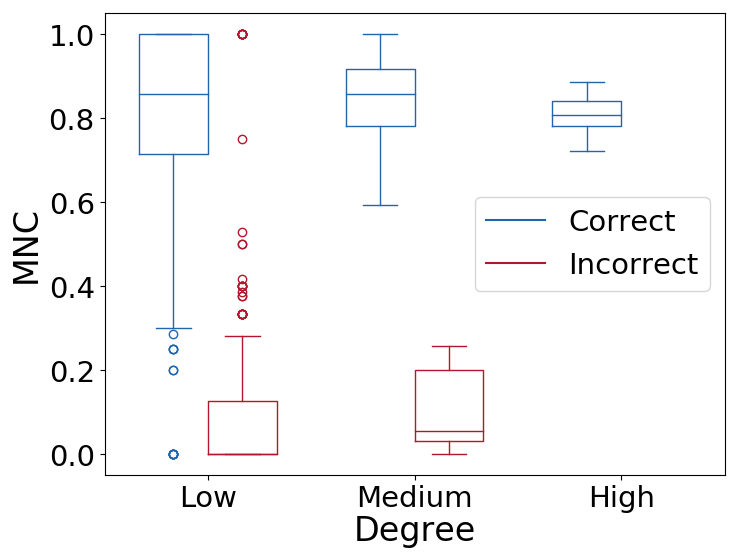}}
    \subfloat[MNC with REGAL\label{fig:easy-distbefore}]{\includegraphics[width=0.22\textwidth]{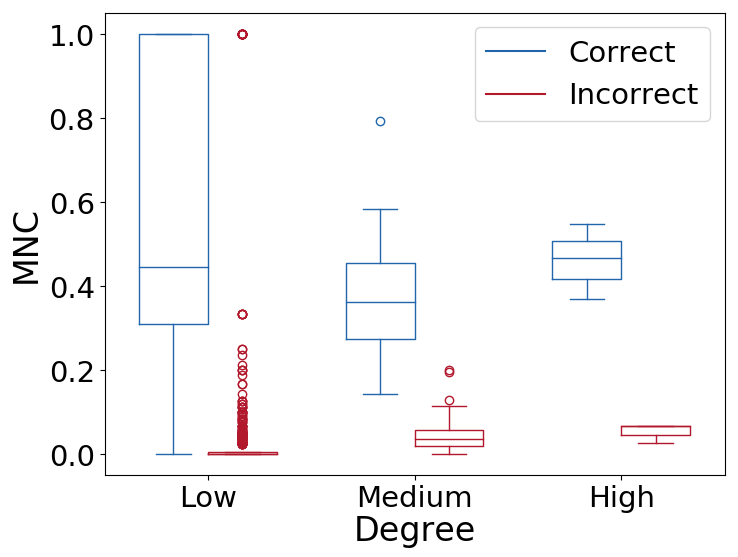}}
    \centering
     \vspace{-0.35cm}
	\caption{MNC of \method and REGAL on the Arenas dataset with 5\% noise.  Compared to REGAL, CONE-Align generates significantly higher MNC for almost all nodes.}
    \label{fig:jaccard}
    \vspace{-0.3cm}
\end{figure}

To further understand MNC, we analyze it on a node-level basis.

\subsubsection{Setup \& Evaluation}
For brevity, we show only REGAL and \method on the Arenas dataset with 5\% noise. 
We split the nodes into three groups by degree: $[0, \frac{\degmax}{3})$, $[\frac{\degmax}{3}, \frac{2\degmax}{3})$, $[\frac{2\degmax}{3}, \degmax]$, where $\degmax$ is the maximum degree, and plot the distribution of MNC for both correctly and incorrectly aligned nodes in Figure~\ref{fig:jaccard}.  



\subsubsection{Results} 
For both methods, 
\textbf{MNC is much higher for correctly aligned nodes} across degree levels, but a few lower degree nodes may be misaligned with high MNC; their smaller neighborhoods may be misaligned together. However, CONE-Align correctly aligns all high-degree nodes.

\section{Conclusion}
\label{sec:conclusion}
\method's success offers the following takeaway: 
the quest for cross-network embedding \emph{comparability} should not neglect intra-network \emph{proximity} information.  
With embedding subspace alignment, we obtain compatibility while capturing proximity.  In the future, this may allow transductive node embeddings to improve other multi-network tasks such as graph classification, where off the shelf they are not applicable~\cite{rgm}. 
Other future work includes using embeddings that model node/edge attributes.

 \section*{Acknowledgements}
{
We would like to thank the reviewers for their constructive feedback.
{This work is supported by}
the NSF under Grant No. IIS 1845491, Army Young Investigator Award No. W911NF1810397, and Adobe, Amazon, and Google faculty awards.
}

\bibliographystyle{plain}
\bibliography{BIB/bibliography}
\end{document}